\documentclass[12pt]{article}
\usepackage{graphicx}
\def\abstracts#1#2#3{{

        \centering{\begin{minipage}{4.5in}\footnotesize\baselineskip=10pt

        \parindent=0pt #1\par

        \parindent=15pt #2\par

        \parindent=15pt #3

        \end{minipage}}\par}}

\begin{document}
\begin{center}
{\large \bf Threshold value of three dimensional \\ bootstrap percolation}\\
\bigskip
Dirk Kurtsiefer\\
{\small Institute for Theoretical Physics, Cologne University, 50937 K\"oln, Germany\\
E-Mail: dk@thp.uni-koeln.de\\}
\end{center}
\bigskip
\abstracts
{The following article deals with the critical value $p_c$ of the
three-dimensional bootstrap percolation. We will check the
behavior of $p_c$ for different lengths of the lattice and
additionally we will scale $p_c$ in the
limit of an infinite lattice.}{}{}
\bigskip
{\bf Introduction}

Bootstrap percolation has been motivated to describe dilute magnetic
systems in which strong competition exists between magnetic and
non-magnetic spins[1]. Kirkpatrick uses bootstrap percolation to
discuss 3D arrays of storage servers and shows that bandwidth is
the critical factor affected by the "fail in place" disorder[2].\\
Let each site of a $d$-dimensional hypercubic lattice be occupied
with probability $p$.
Then we go through each lattice site and look at all $2d$
neighboring places. A lattice site stays occupied if at least
$m$ of the $2d$ adjacent sites are occupied otherwise it becomes
vacant all the time. We use a 0 for vacant and a 1 for occupied site
(see [3] for a detailed description).\\
Let us look at the three-dimensional simple cubic lattice. We define
the threshold as the value $p$ when an infinite cluster is
created. An infinite cluster is here defined as a connected set
of lattice points extending from top to bottom [4].
We get $p_c$ as function of $L$.\\
We calculate the threshold for $m=1,..,5.$ In the case of $m=0$ we
have usual percolation. In the case of $m=6$
lattice sites turn 0 if at least one lattice point is 0.
When we turn down all lattice points at one go we have parallel updating.
When we turn one after another in succession it's sequential updating.
Sequential and parallel updating lead to the same results,
therefore we use the sequential updating to save computing time.
We have used one bit per spin and tested the results in smaller
lattices at one word per spin and odd lattice size. The results
were tested with different random number generators.
Up to 10000 samples were made.\\
\\
{\bf Results}

\begin{figure}[!h]
\centerline{\includegraphics[angle=-90,scale=0.4]{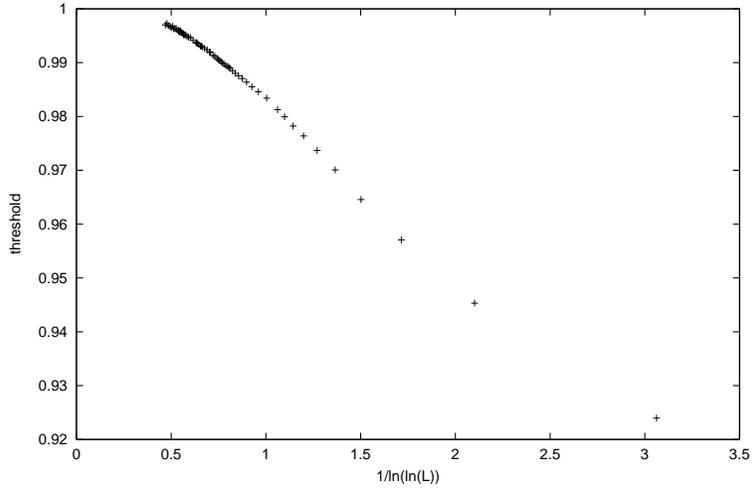}}
\caption{Bootstrap percolation $m=5$, $p_c$ versus $1/\ln (\ln L)$}
\end{figure}

\begin{figure}[!h]
\centerline{\includegraphics[angle=-90,scale=0.4]{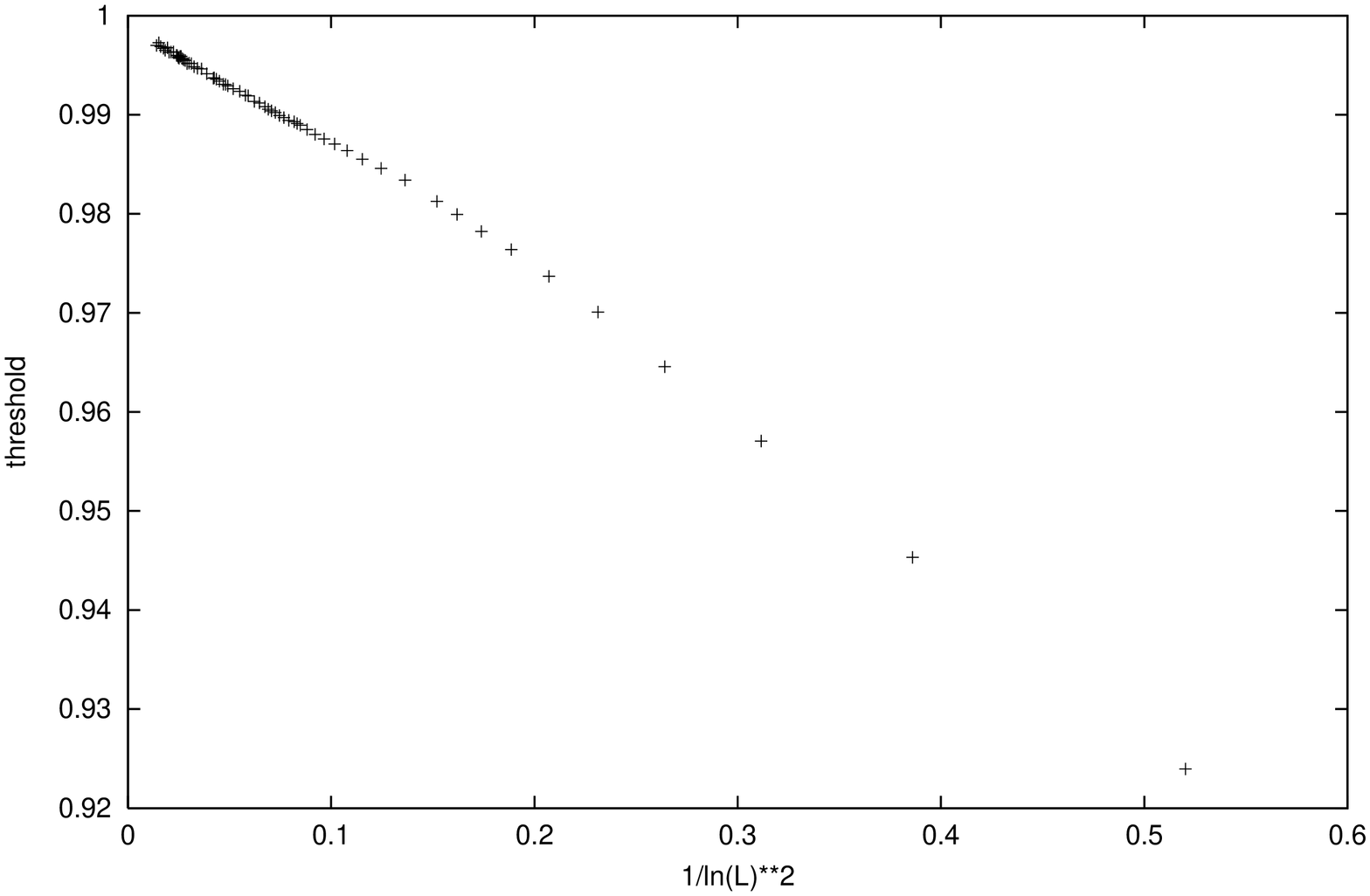}}
\caption {Bootstrap percolation $m=5$, $p_c$ versus $1/(\ln L)^2$}
\end{figure}

\begin{figure}[!h!]
\centerline{\includegraphics[angle=-90,scale=0.4]{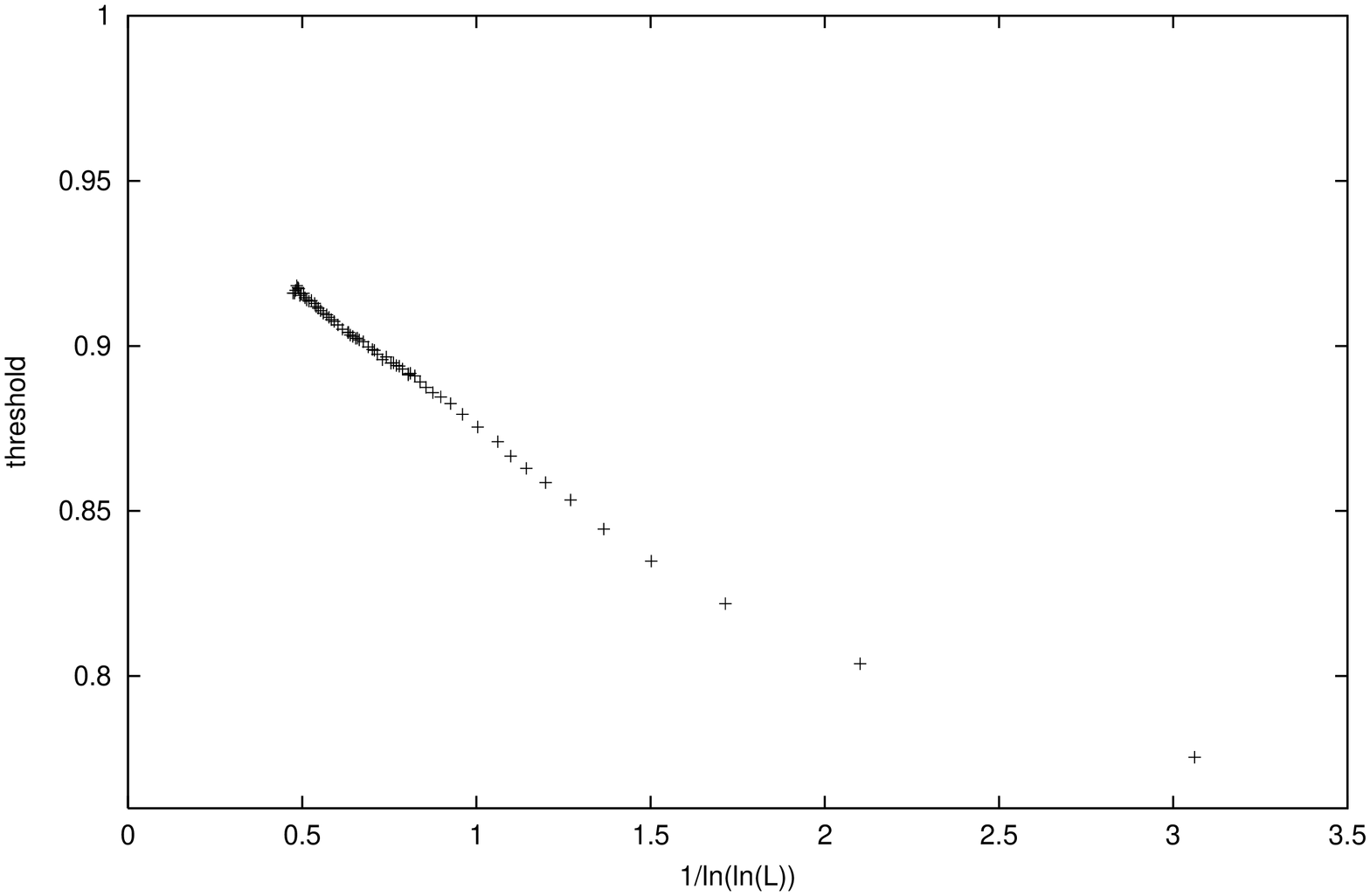}}
\caption{ Bootstrap percolation $m=4$, $p_c$ versus $1/\ln (\ln L)$}
\end{figure}

\begin{figure}[!h!]
\centerline{\includegraphics[angle=-90,scale=0.4]{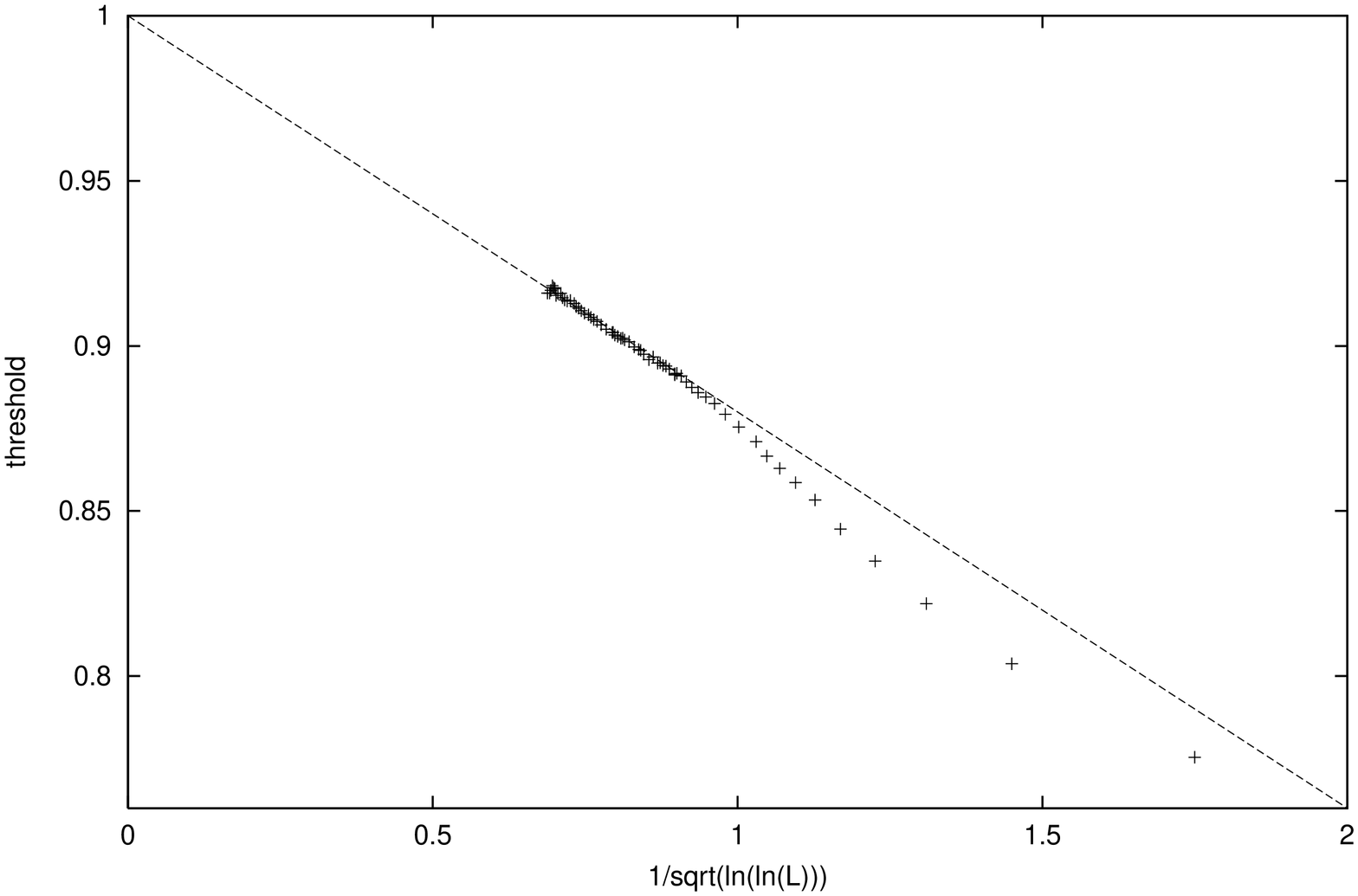}}
\caption{Bootstrap percolation $m=4$, $p_c$ versus $1/\sqrt{\ln (\ln L)}$}
\end{figure}

\begin{figure}[!t]
\centerline{\includegraphics[angle=-90,scale=0.4]{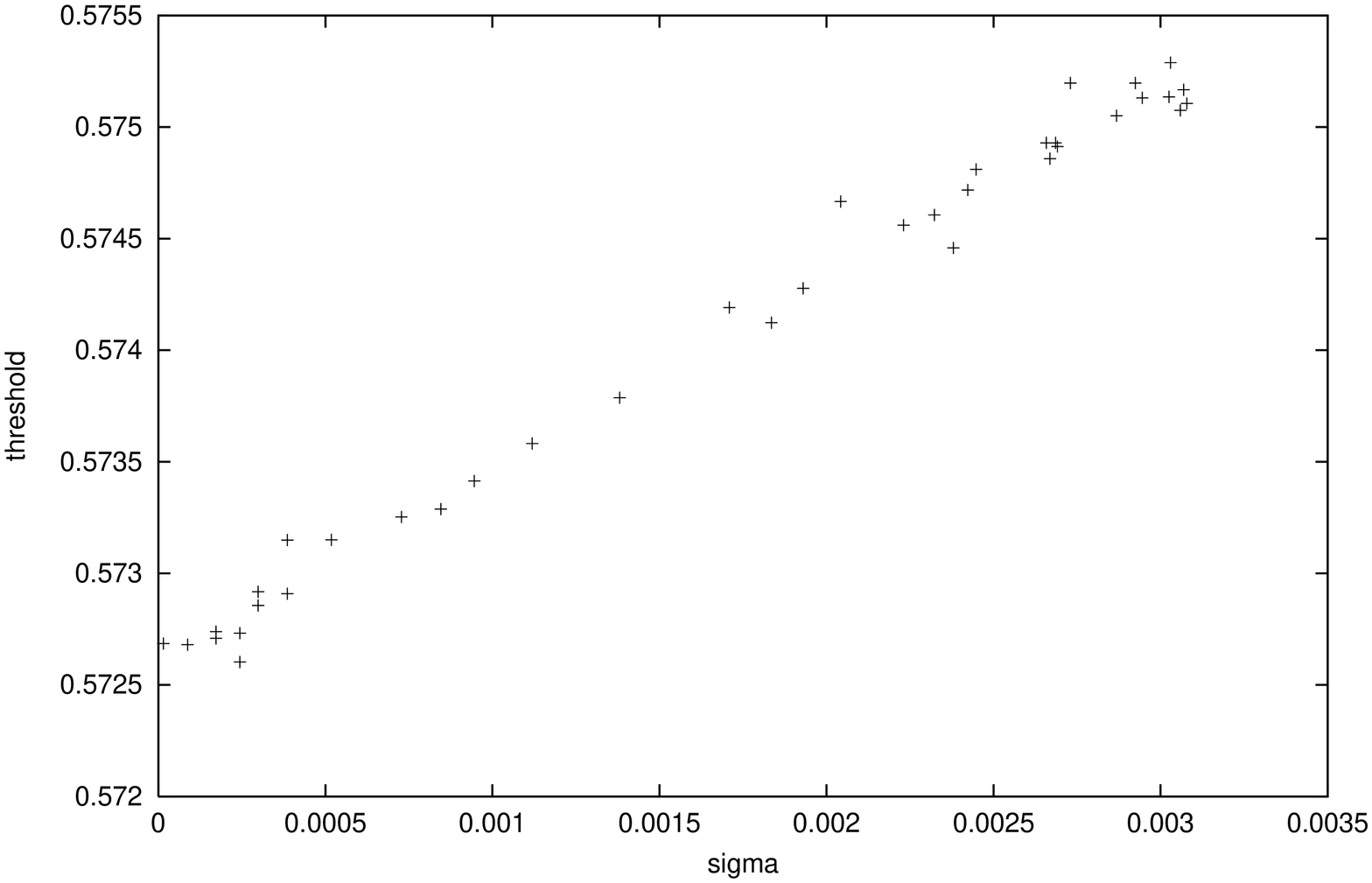}}
\caption{Bootstrap percolation $m=3$, $p_c$ versus $\sigma$}
\end{figure}

\begin{figure}[!t]
\centerline{\includegraphics[angle=-90,scale=0.4]{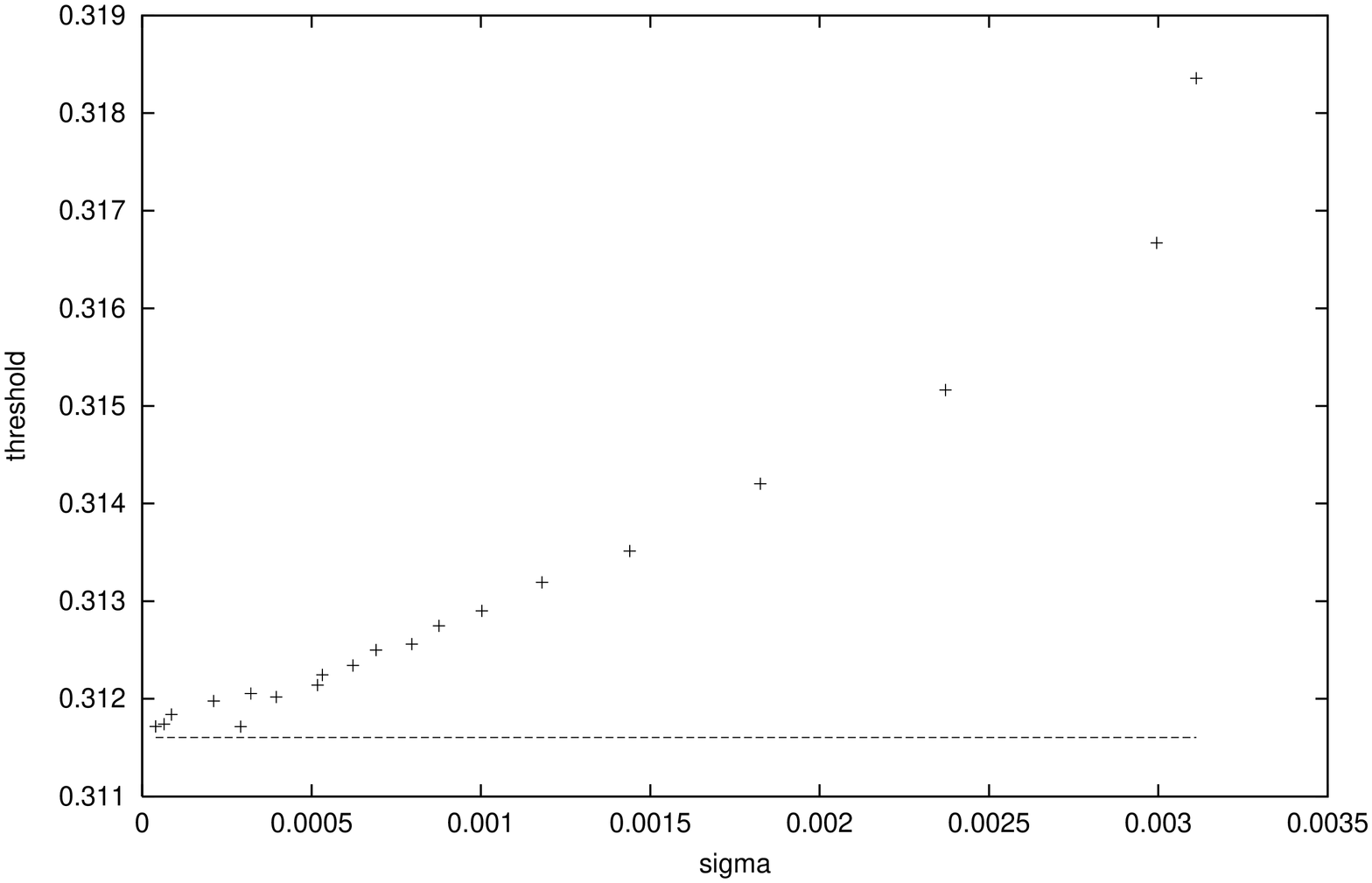}}
\caption{Bootstrap percolation $m=2$, $p_c$ versus $\sigma$}
\end{figure}

\begin{figure}[!t]
\center{\includegraphics[angle=-90,scale=0.4]{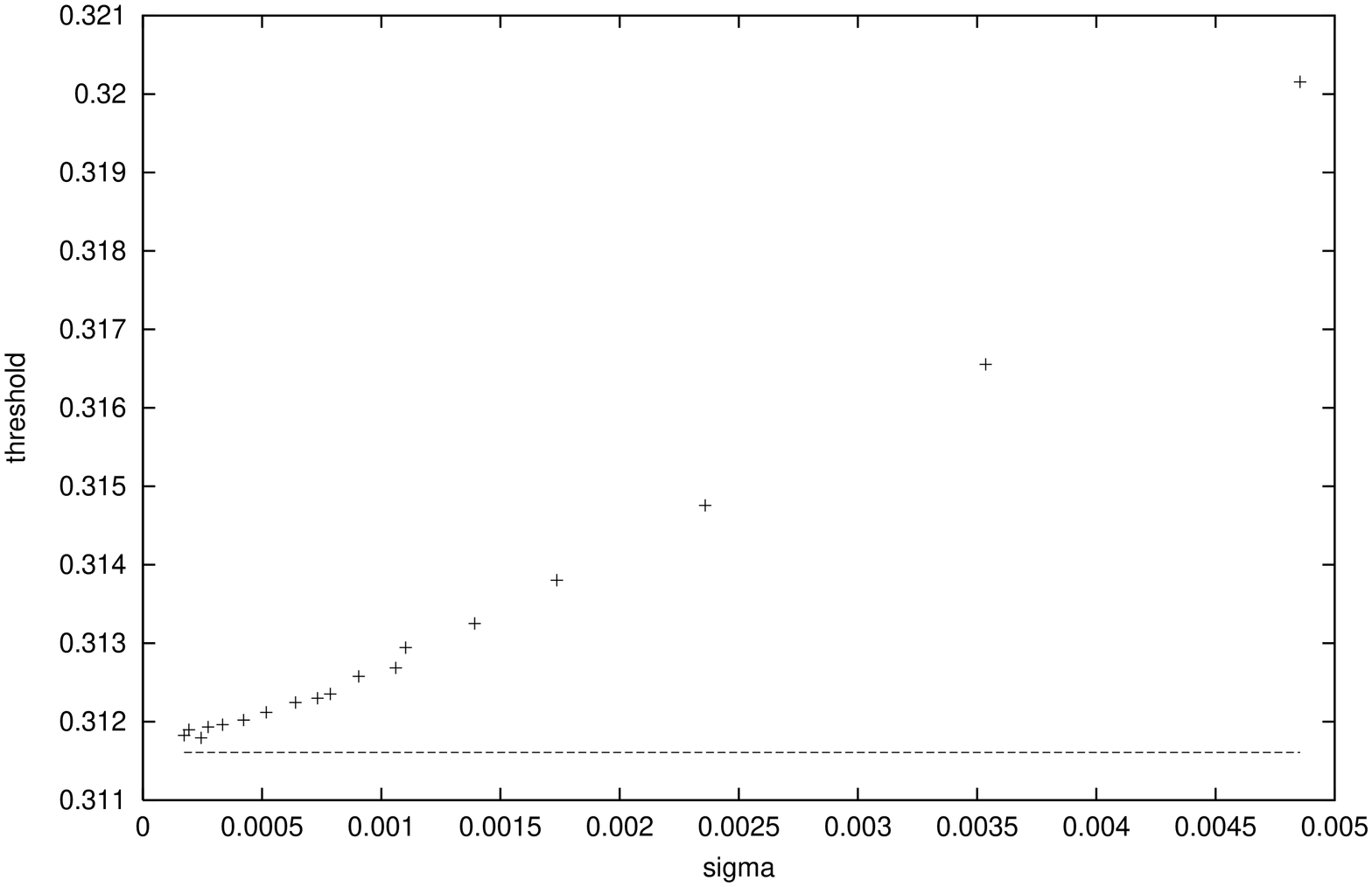}}
\caption{Bootstrap percolation m=1, $p_c$ versus $\sigma$}
\end{figure}

\begin{figure}[!t]
\centerline{\includegraphics[angle=-90,scale=0.4]{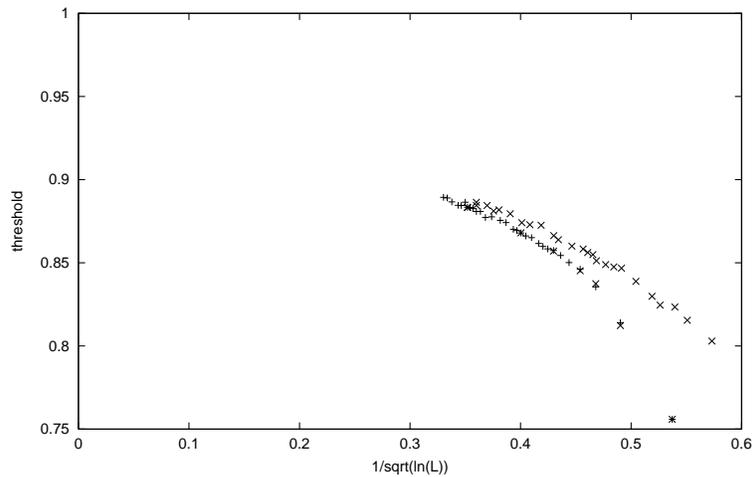}}
\caption{two-dimensional reversible bootstrap percolation $m=3$,
parallel updating, $p_c$ versus $1/\sqrt{\ln L}$, + = one bit per spin
and even lattice size, x = one word per spin with even and odd lattice sizes.
There is a significant difference between odd (upper data) and even
(lower data) lattice sizes.
.}
\end{figure}

\begin{figure}[!t]
\centerline{\includegraphics[angle=-90,scale=0.4]{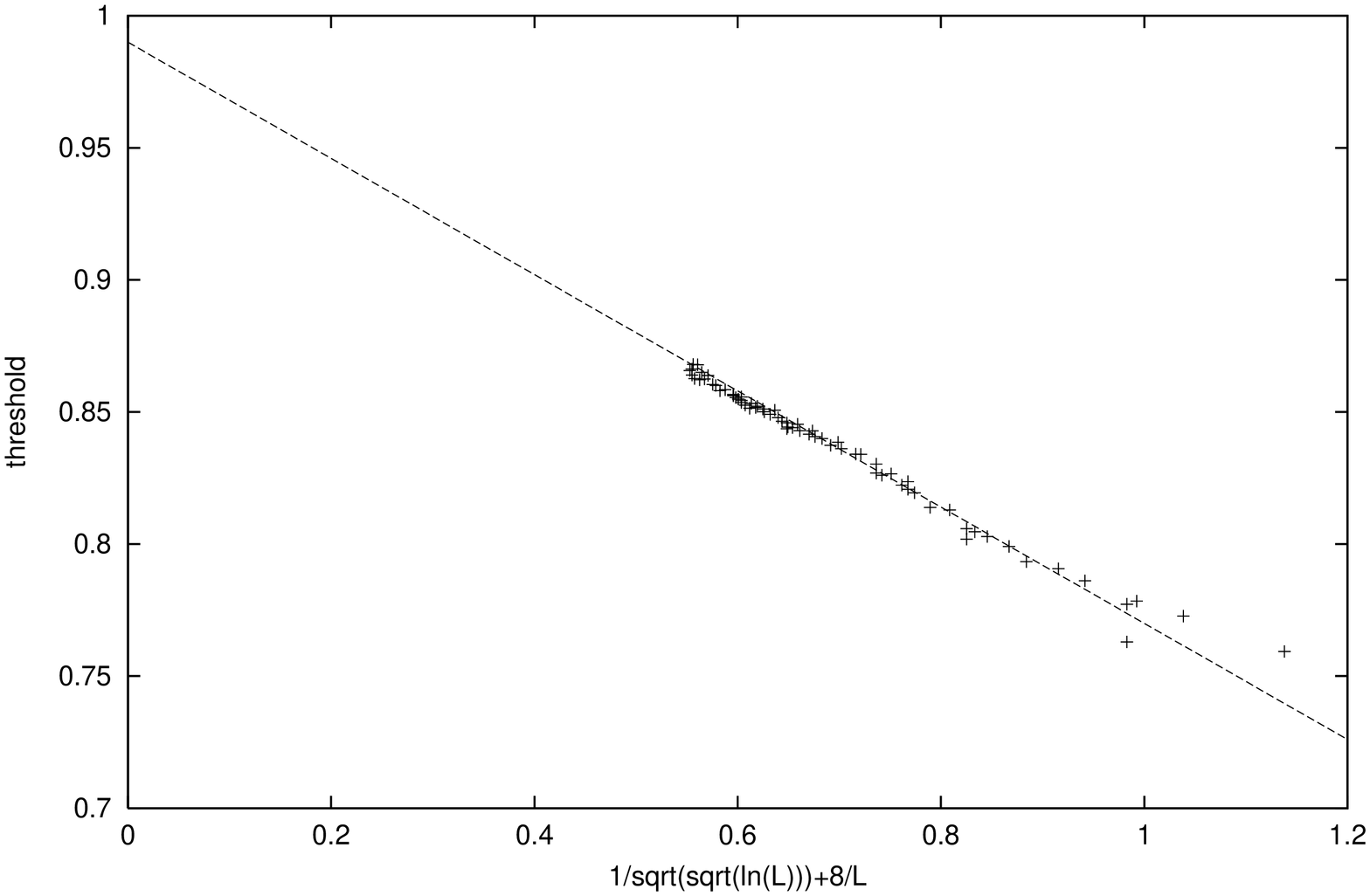}}
\caption{Two-dimensional reversible bootstrap percolation $m=3$,
sequential updating, $p_c$ versus $(\ln L)^{-1/4}+8/L$, $+=1$ bit per Spin
and even lattice size, x $=1$ word per spin with even and odd lattice sizes.}
\end{figure}

In the case of $m=5$ and $m=4$ all the lattice sites become to 0
if there is no infinite cluster. Schonmann shows rigorously[5]
that for all $m>d$ the threshold for $L\rightarrow\infty$ is 1.
Besides it was proven[6] that for $m=4$, $1-p_c$ scales as $1/\ln (\ln L)$.
In agreement with earlier results[7] we obtained fig.1 to 4.
In the case of $m=5$ we have a straight line leading to 1 if we plot
$p_c$ versus $1/(\ln L)^2$ (fig.2). The theoretical
prediction for $m=4$ could not be confirmed by
simulations (fig.3). The best straight line $L\rightarrow\infty$
is found for $p_c$ versus $1/\sqrt{\ln (\ln L)}$ (fig.4).
We have simulated lattices from $L=4$ up to $L=4544$.
The biggest lattice sizes were simulated on the Cray T3E
parallel computer in HLRZ J\"ulich on up to 128 processors.\\
The case $m=3$ is the only one that has a threshold for
$L \rightarrow \infty$ between 1 and the normal threshold
0.3116[4]. When we count the occupied sites we can observe a inflection
point for the fraction $\pi$ of finally occupied sites
as a function of the initial concentration $p$. But the maximal value
of $d\pi/dp$ is reached at about $p=0.62$ for all sizes of $L$.
Therefore the method doesn't lead to a peak at the transition point
or doesn't have a sufficiently sharp peak to useful.
We must use instead the Hoshen-Kopelman algorithm[4]
for cluster-analysis to find $p_c$. Hoshen-Kopelman have made
a procedure that counts all cluster sizes[8].
When $k_i$ is the number of
clusters with $i$ sites we get $\chi$ as
$\sum_i^\prime  i^2 k_i$. We get $p_{av}$ from the maximum of $\chi$, and
average over many samples. We know that $p_{av}-p_c \propto\sigma$ and
$\sigma\propto L^{-\nu}$ with $\sigma$ the width
$\sigma=\sqrt{(<p_{av}^2>-<p_{av}>^2)}$. $p_{av}$ is the threshold for any
$L<\infty$ and $p_c$ is the threshold for infinite $L$. We plot
$p_{av}$ versus $\sigma$ and in agreement with [9] we get a threshold
of 0.5726 $\pm$ 0.0001 (fig.5). We have a slight difference
between even and uneven $L$. The Hoshen-Kopelman
algorithm prolongs computation time to about 2.2 times it's value
without Hoshen-Kopelman.
But computer memory limited our simulations because the larger
lattice sizes need up to 1 GByte storage.
The largest lattice size we have simulated is $L=1696$, the smallest
one was $L=31$.\\
The cases $m=2$ and $m=1$ results in a value for the threshold
similarly to the one expected, i.e. 0.3116 (fig. 6,7).
Here the biggest lattice size is $L=1504$ at $m=2$ and $L=1120$ at
$m=1$.\\
We have plotted only the even lattice sizes because when the
lattice size is odd the width gets larger. The reason for
getting to larger $L$ at $m=3$ is, we can save memory by
Hoshen-Kopelman because the cluster becomes
rectangular.\\
Finally, let us have a view at the two-dimension reversible bootstrap. It also
called biased majority rule model, because the state $x$ at time
$t+1$ is a majority function of the four neighbours at time $t$ with a
bias toward occupation in case of a tie. Schonmann shows
mathematically that for $m=3$ is $C_1/\sqrt{\ln L} \leq
1-p(L,\alpha) \leq C_2/\sqrt{\ln L}$[10,11]. When we have
parallel updating this may be correct[12] but the differences between
even and odd lattice size $L$ are significant (fig.8). When we apply
sequential updating that doesn't make any difference, but the
critical point is lower. Sequential
updating gets more empty sites by the same $p$ so we obtained a
lower threshold. When we plot $p_c$ versus
$(\ln L)^{-1/4}+8/L$ (fig.9) we get a line
extrapolated to 1.\\
\\
{\bf Summary}

We have independently confirmed older results even when they
disagreed somewhat with theory for $m=4$ in three dimensions,
and $m=3$ in two. Of course, the theories are valid only
for $L \rightarrow \infty$ and perhaps not for our data.
Our results are more accurate because we simulate bigger lattices
and more lattice sizes. We get the following empirical finite-size corrections:
\begin{center}
\begin{tabular}{|c|c|c|c|}
\hline
\multicolumn{4}{|c|}{critical behavior $p_c$ of the} \\
\multicolumn{4}{|c|}{three-dimensional bootstrap percolation}\\
\hline\hline
$m$ & $p_c$ for $L \to \infty$ & versus & rigorously\\
\hline
5 & 1 & $1/(\ln L)^2$ & \\
\hline
4 & 1 & $1/\sqrt{\ln (\ln L)}$ & $1/\ln (\ln L)$\\
\hline
3 & 0.5726 $\pm$ 0.0001& $\sigma $ &  \\
\hline
2 & 0.3116 & $\sigma$ & 0.3116\\
\hline
1 & 0.3116 & $\sigma$ & 0.3116\\
\hline
0 & 0.3116 & $\sigma$ & 0.3116\\
\hline
\hline
\hline
\multicolumn{4}{|c|}{two-dimensional bootstrap percolation}\\
\hline
\hline
3 & 1 & $(\ln L)^{-1/4}$ & $(\ln L)^{-1/2}$\\
\hline
\end{tabular}
\end{center}
\bigskip
{\bf Acknowledgments}

Thanks to Scott Kirkpatrick for inspiration to this article, Dietrich
Stauffer for suggestions, Joan Adler and Nilton Branco for improving
the text and GIF for indirect support.\\
\\
{\bf  References}
\begin{small}
\begin{enumerate}
\item J.Chalupa, P.L.Leath and G.R.Reich, J.Phys.C 12 (1979) L31
\item S.Kirkpatrick, appear in Physica A,2002
\item J.Adler, Physica A 171 (1991) 453
\item D.Stauffer and A.Aharony, Introduction to Percolation
Theory, (Taylor and Francis, London,1992)
\item R.H.Schonmann, Ann Probab 20 (1992) 174
\item R.Cerf and E.Cirildo, Ann Probab 27 (1999) 1837
\item S.S.Manna and D.Stauffer, Physica A 162 (1989) 20
\item J.Hoshen R.Kopelman, Physical Review B 14 (1976) 3438
\item N.S.Branco and C.J.Silva, Int J Mod Phys C 10 (1999) 921
\item R.H.Schonmann, Journal of Statistical Physics 58 (1990) 1239
\item R.H.Schonmann, Physica A 167 (1990) 619
\item D.Stauffer, Physica Scripta T35 (1991) 66
\end{enumerate}
\end{small}
\end{document}